\documentclass[11pt]{article}%
\usepackage{amsfonts}
\usepackage{amsmath}
\usepackage{amssymb}
\usepackage{graphicx}%
\providecommand{\U}[1]{\protect\rule{.1in}{.1in}}
\newtheorem{theorem}{Theorem}
\newtheorem{acknowledgement}[theorem]{Acknowledgement}

\newtheorem{lemma}[theorem]{Lemma}

\begin{document}

\title{On the Wigner Distribution of the Reduced Density Matrix}
\author{Maurice A. de Gosson\thanks{maurice.de.gosson@univie.ac.at (corresponding
author)}\\University of Vienna\\Faculty of Mathematics (NuHAG)
\and Charlyne de Gosson\\University of Vienna\\Faculty of Mathematics (NuHAG)}
\maketitle

\begin{abstract}
Consider a bipartite quantum system consisting of two subsystems $A$ and $B$.
The reduced density matrix of$A$ a is obtained by taking the partial trace
with respect to $B$. In this work we will show that the Wigner distribution of
this reduced density matrix is obtained by integrating the total Wigner
distribution with respect to the phase space variables corresponding to the
subsystem $B$. The proof we give is rigorous (as opposed to those found in the
literature) and makes use of the Weyl--Wigner--Moyal phase space formalism.
Our main result is applied to general Gaussian mixed states, of which it gives
a particularly simple and precise description. We also briefly discuss the
purification of a mixed from the Wigner formalism point of view.

\end{abstract}

\section{Introduction}

A quantum state is abstractly defined as a non-negative operator with trace
one on some Hilbert space \cite{blabru,Birkbis,Quanta,QHA}. In this paper we
will assume that this Hilbert space is $L^{2}(\mathbb{R}^{n})$ (the square
integrable functions). We will deal with a bipartite system consisting of two
parts $A$ and $B$ with respective Hilbert spaces $L^{2}(\mathbb{R}^{n_{A}})$
and $L^{2}(\mathbb{R}^{n_{B}})$ (such systems play a central role in the study
of quantum entanglement). We denote by $\mathcal{L}_{1}(L^{2}(\mathbb{R}%
^{n}))$ (resp. $\mathcal{L}_{1}(L^{2}(\mathbb{R}^{n_{A}}))$) the spaces of
trace-class operators on $L^{2}(\mathbb{R}^{n})$ (\textit{resp}. on
$L^{2}(\mathbb{R}^{n_{A}})$). Setting $n=n_{A}+n_{B}$ there exists
\cite{Nielsen} a unique mapping
\begin{gather*}
\operatorname*{Tr}\nolimits^{B}:\mathcal{L}_{1}(L^{2}(\mathbb{R}%
^{n}))\longrightarrow\mathcal{L}_{1}(L^{2}(\mathbb{R}^{n_{A}}))\\
\widehat{\rho}\longmapsto\widehat{\rho}^{A}=\operatorname*{Tr}\nolimits^{B}%
(\widehat{\rho})
\end{gather*}
such that for every $\widehat{Q}^{A}\in\mathcal{B}(L^{2}(\mathbb{R}^{n_{A}}))$
(the bounded operators) the identity
\begin{equation}
\operatorname*{Tr}(\widehat{\rho}(\widehat{Q}^{A}\otimes\widehat{I}%
^{B}))=\operatorname*{Tr}(\widehat{\rho}^{A}\widehat{Q}^{A}) \label{parttrace}%
\end{equation}
holds for all $\widehat{\rho}\in\mathcal{L}_{1}(L^{2}(\mathbb{R}^{n}))$ where
$\widehat{I}^{B}$ is the identity on $L^{2}(\mathbb{R}^{n_{B}})$ (the
left-hand side of (\ref{parttrace}) is well defined since $\mathcal{L}%
_{1}(L^{2}(\mathbb{R}^{n}))$ is a two-sided ideal in $\mathcal{B}%
(L^{2}(\mathbb{R}^{n}))$). The map $\operatorname*{Tr}^{B}$ is called
\textquotedblleft partial trace with respect to $B$\textquotedblright. Of
particular interest is the case where $\widehat{\rho}$ has trace one and is
positive semi-definite: $\widehat{\rho}\geq0$, because then $\widehat{\rho}$
is a quantum density matrix and the operator $\widehat{\rho}^{A}%
=\operatorname*{Tr}^{B}(\widehat{\rho})$ is the reduced density matrix on
$L^{2}(\mathbb{R}^{n_{A}})$.

The aim of this paper is to show that the reduced density matrix
$\widehat{\rho}^{A}$ can be determined in a very easy and explicit way if one
applies the Weyl--Wigner--Moyal phase space formalism \cite{Moyal,Weyl} to the
Wigner distribution \cite{Neumann} of the full density matrix $\widehat{\rho}%
$. In fact, we are going to prove the following result, where we use the
notation $z=(z_{A},z_{B})$, $z_{A}=(x_{A},p_{A})$, $z_{B}=(x_{B},p_{B})$ for
the phase space variables:

\begin{theorem}
\label{Thm1}Let $\widehat{\rho}\in\mathcal{L}_{1}(L^{2}(\mathbb{R}^{n}))$ be a
density operator and
\begin{equation}
\rho(z)=\left(  \tfrac{1}{2\pi\hbar}\right)  ^{n}\int_{\mathbb{R}^{n}%
}e^{-\frac{i}{\hbar}py}\langle x+\tfrac{1}{2}y|\widehat{\rho}|x-\tfrac{1}%
{2}y\rangle dy \label{Wig}%
\end{equation}
its Wigner distribution ($x=(x_{A},x_{B})$). If $\rho$ decreases sufficiently
fast at infinity (as described in Lemma \ref{Lemma1}) then the Wigner
distribution of the reduced density matrix $\widehat{\rho}^{A}%
=\operatorname*{Tr}^{B}(\widehat{\rho})$ is
\begin{equation}
\rho^{A}(z_{A})=\int_{\mathbb{R}^{2n_{B}}}\rho(z_{A},z_{B})dz_{B}. \label{F1}%
\end{equation}

\end{theorem}

As a simple illustration we will apply this result to Gaussian mixed states.
We also consider the purification of a density matrix $\rho^{A}\in
\mathcal{L}_{1}(L^{2}(\mathbb{R}^{n_{A}}))$; this will lead us to a new
property of the Wigner function. Namely, we show that for every normalized
$\psi\in L^{2}(\mathbb{R}^{n})$ there exists an orthonormal basis $(\phi
_{j}^{A})$ of $L^{2}(\mathbb{R}^{n_{A}})$ and numbers $\lambda_{j}\geq0$,
$\sum_{j}\lambda_{j}=1$, such that
\begin{equation}
\int_{\mathbb{R}^{2n_{B}}}W\psi(z_{A},z_{B})dz_{B}=\sum_{j}\lambda_{j}%
W^{A}\phi_{j}^{A}(z_{A})~. \label{F2}%
\end{equation}

\section{Preliminaries}

A density operator $\widehat{\rho}$ on $L^{2}(\mathbb{R}^{n})$ is a compact
self-adjoint operator hence the spectral theorem ensures us of the existence
of an orthonormal set of vectors $(\psi_{j})$ in $L^{2}(\mathbb{R}^{n})$ and
of positive constants $\lambda_{j}$ summing up to one such that
\begin{equation}
\widehat{\rho}=\sum_{j}\lambda_{j}|\psi_{j}\rangle\langle\psi_{j}|~.
\label{sum}%
\end{equation}
Since for every index $k$ we have $\widehat{\rho}|\psi_{k}\rangle=\lambda
_{k}|\psi_{k}\rangle$ the $\lambda_{j}$ are the eigenvalues of $\widehat{\rho
}$ and the $|\psi_{j}\rangle$ the corresponding eigenvectors. Inserting this
sum into definition (\ref{Wig}) of the Wigner distribution we get
\begin{equation}
\rho(x,p)=\left(  \tfrac{1}{2\pi\hbar}\right)  ^{n}\sum_{j}\lambda_{j}%
\int_{\mathbb{R}^{n}}e^{-\frac{i}{\hbar}py}\langle x+\tfrac{1}{2}y|\psi
_{j}\rangle\langle\psi_{j}|x-\tfrac{1}{2}y\rangle dy
\end{equation}
(we are using in this formula the standard splitting $x=(x_{A},x_{B})$,
$p=(p_{A},p_{B})$). Since we have
\[
\langle x+\tfrac{1}{2}y|\psi_{j}\rangle=\psi_{j}(x+\tfrac{1}{2}y)\text{
},\text{ }\langle\psi_{j}|x-\tfrac{1}{2}y\rangle=\psi_{j}^{\ast}(x-\tfrac
{1}{2}y),
\]
it follows that%
\[
\left(  \tfrac{1}{2\pi\hbar}\right)  ^{n}\int_{\mathbb{R}^{n}}e^{-\frac
{i}{\hbar}py}\langle x+\tfrac{1}{2}y|\psi_{j}\rangle\langle\psi_{j}%
|x-\tfrac{1}{2}y\rangle dy=W\psi_{j}(x,p)
\]
where
\begin{equation}
W\psi_{j}(x,p)=\left(  \tfrac{1}{2\pi\hbar}\right)  ^{n}\int_{\mathbb{R}^{n}%
}e^{-\frac{i}{\hbar}py}\psi_{j}(x+\tfrac{1}{2}y)\psi_{j}^{\ast}(x-\tfrac{1}%
{2}y)dy \label{wigdef}%
\end{equation}
is the usual Wigner transform \cite{gowig,Wigner} of $\psi_{j}$. The Wigner
distribution of the density matrix $\widehat{\rho}$ is thus a convex sum of
Wigner transforms \cite{Quanta}:
\[
\rho(x,p)=\sum_{j}\lambda_{j}W\psi_{j}(x,p).
\]

Let now $W(\phi,\psi)$ be the cross-Wigner transform \cite{Wigner} of
$\phi,\psi\in L^{2}(\mathbb{R}^{n})$. By definition \cite{gowig}%
\begin{equation}
W(\phi,\psi)(x,p)=\left(  \tfrac{1}{2\pi\hbar}\right)  ^{n}\int_{\mathbb{R}%
^{n}}e^{-\frac{i}{\hbar}py}\phi(x+\tfrac{1}{2}y)\psi^{\ast}(x-\tfrac{1}{2}y)dy
\label{crossWig}%
\end{equation}
hence, in particular, $W(\psi,\psi)=W\psi$. We recall the following classical
formula from Weyl calculus \cite{Birkbis,gowig}: if $\widehat{Q}$ is a bounded
operator on $L^{2}(\mathbb{R}^{n})$ then, setting $z=(x,p)$,%
\begin{equation}
\langle\phi|\widehat{Q}\psi\rangle=\int_{\mathbb{R}^{2n}}q(z)W(\phi,\psi)(z)dz
\label{symbol1}%
\end{equation}
where $q(z)$ is the Weyl symbol of $\widehat{Q}$, that is,
\begin{equation}
q(x,p)=\int_{\mathbb{R}^{n}}e^{-\frac{i}{\hbar}py}\langle x+\tfrac{1}%
{2}y|\widehat{Q}|x-\tfrac{1}{2}y\rangle dy~. \label{symbol2}%
\end{equation}

We will use he notation $\widehat{Q}=\operatorname*{Op}\nolimits_{\mathrm{W}%
}(q)$ to denote the Weyl operator with symbol $q$; formally
\begin{equation}
\operatorname*{Op}\nolimits_{\mathrm{W}}(q)\psi(x)=\left(  \tfrac{1}{2\pi
\hbar}\right)  ^{n}\int_{\mathbb{R}^{2n}}e^{\frac{i}{\hbar}p(x-yy}q(\tfrac
{1}{2}(x+y),p)\psi(y)dpdy. \label{opw1}%
\end{equation}
Comparison with formula (\ref{Wig}) shows that the Wigner distribution of a
density matrix is just the Weyl symbol of that density matrix divided by
$(2\pi\hbar)^{n}$, that is,
\begin{equation}
\widehat{\rho}=(2\pi\hbar)^{n}\operatorname*{Op}\nolimits_{\mathrm{W}}(\rho)~.
\label{rhoweyl}%
\end{equation}

We will use the following criterion due to Shubin (\cite{sh87}, p. 203) for
determining whether a given function on phase space is the Weyl symbol of an operator:

\begin{lemma}
\label{Lemma1}Assume that $q\in C^{\infty}(\mathbb{R}^{2n})$ and that there
exist for every multi-index $\alpha\in\mathbb{N}^{n}$ a constant $C_{\alpha
}>0$ such that
\begin{equation}
|\partial_{z}^{\alpha}q(z)|\leq C_{\alpha}(1+|z|^{2})^{(m-|\alpha|)/2}.
\label{Shubin}%
\end{equation}
If $m<-2n$, then $\widehat{Q}=\operatorname*{Op}_{\mathrm{W}}(q)$ is of trace
class and we have%
\begin{equation}
\operatorname*{Tr}(\widehat{Q})=\left(  \tfrac{1}{2\pi\hbar}\right)  ^{n}%
\int_{\mathbb{R}^{2n}}q(z)dz \label{trace}%
\end{equation}
where the integral is absolutely convergent.
\end{lemma}

In particular, if $\rho$ satisfies the conditions above then the operator
$\widehat{\rho}=(2\pi\hbar)^{n}\operatorname*{Op}\nolimits_{\mathrm{W}}(\rho)$
is of trace class and its trace is given by the formula%
\begin{equation}
\operatorname*{Tr}(\widehat{\rho})=\int_{\mathbb{R}^{2n}}\rho(z)dz
\label{integtrace}%
\end{equation}
(this formula is often sloppily used in the literature, but one should be
aware of the fact that it only holds under rather stringent conditions on
$\rho$; see our discussion and the references in \cite{Quanta,Birkbis}).

\section{Proof of Theorem \ref{Thm1}}

We will again use the notation $z_{A}=(x_{A},p_{A})$, $z_{B}=(x_{B},p_{B})$,
and $z=(z_{A},z_{B})$.

The partial trace operator (\ref{parttrace}) can be explicitly calculated
\cite{Nielsen} using the formula
\begin{equation}
\langle\phi^{A}|\widehat{\rho}^{A}\psi^{A}\rangle_{_{A}}=\sum_{j}\langle
\phi^{A}\otimes\phi_{j}^{B}|\widehat{\rho}(\psi^{A}\otimes\phi_{j}^{B}%
)\rangle\label{parttrace1}%
\end{equation}
where $\psi^{A},\phi^{A}\in L^{2}(\mathbb{R}^{n_{A}})$ and $(\phi_{j}^{B})$ is
an arbitrary orthonormal basis of $L^{2}(\mathbb{R}^{n_{B}})$; $\langle
\cdot|\cdot\rangle_{_{A}}$ is the bra-ket on $L^{2}(\mathbb{R}^{n_{A}})$.
Taking into account the relations $\widehat{\rho}^{A}=(2\pi\hbar)^{n_{A}%
}\operatorname*{Op}_{\mathrm{W}}(\rho^{A})$ and $\widehat{\rho}=(2\pi
\hbar)^{n}\operatorname*{Op}_{\mathrm{W}}(\rho)$ we have, in view of formula
(\ref{symbol1}),
\begin{gather}
\langle\phi^{A}|\widehat{\rho}^{A}\psi^{A}\rangle_{_{A}}=(2\pi\hbar)^{n_{A}%
}\int_{\mathbb{R}^{2n_{A}}}\rho^{A}(z_{A})W^{A}(\phi^{A},\psi^{A}%
)(z_{A})dz_{A}\label{calcul1}\\
\langle\phi^{A}\otimes\phi_{j}^{B}|\widehat{\rho}(\psi^{A}\otimes\phi_{j}%
^{B})\rangle=(2\pi\hbar)^{n}\int_{\mathbb{R}^{2n}}\rho(z)W(\phi^{A}\otimes
\phi_{j}^{B},\psi^{A}\otimes\phi_{j}^{B})(z)dz \label{calcul2}%
\end{gather}
where $W^{A}(\phi^{A},\psi^{A})$ is the cross-Wigner transform on
$L^{2}(\mathbb{R}^{n_{A}})$. In view of the obvious tensor product property%
\[
W(\phi^{A}\otimes\phi_{j}^{B},\psi^{A}\otimes\phi_{j}^{B})=W^{A}(\phi^{A}%
,\psi^{A})\otimes W^{B}\phi_{j}^{B}%
\]
the identity (\ref{calcul2}) becomes%
\begin{multline*}
\langle\phi^{A}\otimes\phi_{j}^{B}|\widehat{\rho}(\psi^{A}\otimes\phi_{j}%
^{B})\rangle=(2\pi\hbar)^{n}\int_{\mathbb{R}^{2n_{A}}}\left[  \int%
_{\mathbb{R}^{2n_{B}}}\rho(z)W^{B}\phi_{j}^{B}(z_{B})dz_{B}\right] \\
\times W^{A}(\phi^{A},\psi^{A})(z_{A})dz_{A}%
\end{multline*}
and comparison with the equality (\ref{calcul1}) shows that we have%
\[
(2\pi\hbar)^{n_{A}}\rho^{A}(z_{A})=(2\pi\hbar)^{n}\sum_{j}\int_{\mathbb{R}%
^{2n_{B}}}\rho(z_{A},z_{B})W^{B}\phi_{j}^{B}(z_{B})dz_{B}~,
\]
that is, since $n=n_{A}+n_{B}$,%
\begin{equation}
\rho^{A}(z_{A})=(2\pi\hbar)^{n_{B}}\sum_{j}\int_{\mathbb{R}^{2n_{B}}}%
\rho(z_{A},z_{B})W^{B}\phi_{j}^{B}(z_{B})dz_{B}~. \label{rhoza}%
\end{equation}
There remains to prove that the right-hand side of this equality is just the
integral $\int_{\mathbb{R}^{2n_{B}}}\rho(z_{A},z_{B})dz_{B}$. Let us freeze
the value of $z_{A}$ and set $f(z_{B})=\rho(z_{A},z_{B})$. Since we are
assuming that $\rho$ satisfies the Shubin estimates (\ref{Shubin}) we have,
for some $m<-n/2$,
\begin{align*}
|\partial_{z_{B}}^{\beta}f(z_{B})|  &  \leq C_{\beta}(1+|z_{A}|^{2}%
+|z_{B}|^{2})^{m-|\beta|/2}\\
&  \leq C_{\beta}(1+|z_{B}|^{2})^{m-|\beta|/2}.
\end{align*}
The inequality $m<-n/2$ \textit{de facto} implies that $m<-n_{B}/2$ hence it
follows from Lemma \ref{Lemma1} that the operator $\widehat{F}%
=\operatorname*{Op}_{\mathrm{W}}(f)$ is of trace class on $L^{2}%
(\mathbb{R}^{2n_{B}})$ and that its trace is the absolutely convergent
integral%
\begin{equation}
\operatorname*{Tr}(\widehat{F})=\left(  \tfrac{1}{2\pi\hbar}\right)  ^{n_{B}%
}\int_{\mathbb{R}^{2n_{B}}}f(z_{B})dz_{B}~. \label{trf}%
\end{equation}
Returning to the identity (\ref{rhoza}) we have, for every $z_{A}$,
\begin{align*}
\rho^{A}(z_{A})  &  =\sum_{j}\int_{\mathbb{R}^{2n_{B}}}f(z_{B})W^{B}\phi
_{j}^{B}(z_{B})dz_{B}\\
&  =\sum_{j}\langle\phi_{j}^{B}|\widehat{F}\phi_{j}^{B}\rangle\\
&  =\operatorname*{Tr}(\widehat{F})
\end{align*}
the last identity being just the algebraic definition of the trace of a
positive semi-definite trace class operator \cite{blabru,Birkbis} in terms of
an (arbitrary) orthonormal basis $(\phi_{j}^{B})$. Taking (\ref{trf}),
(\ref{rhoza}), and the definition of the function $f$ into account, we get
\[
\rho^{A}(z_{A})=\int_{\mathbb{R}^{2n_{B}}}\rho(z_{A},z_{B})dz_{B}%
\]
which was to be proven.

\section{Application to Gaussian states}

A (centered) Gaussian quantum state has Wigner distribution%
\begin{equation}
\rho(z)=\frac{1}{(2\pi)^{n}\sqrt{\det\Sigma}}e^{-\frac{1}{2}z^{T}\Sigma^{-1}z}
\label{rhoG}%
\end{equation}
where the covariance matrix $\Sigma$ is symmetric positive definite and of
dimension $2n\times2n$ (we will write for short $\Sigma>0$). The uncertainty
principle imposes that, in addition, $\Sigma$ must satisfies the condition
\cite{Birkbis,Narconnell,sisumu}
\[
\Sigma+\frac{i\hbar}{2}J\geq0
\]
where $J$ is the standard symplectic matrix in the phase space splitting
$z=(z_{A},z_{B})$, that is, $J=J_{A}\oplus J_{B}$ with
\[
J_{A}=%
\begin{pmatrix}
0_{n_{A}\times n_{A}} & I_{n_{A}\times n_{A}}\\
-I_{n_{A}\times n_{A}} & I_{n_{A}\times n_{A}}%
\end{pmatrix}
\text{ },\text{ }J_{B}=%
\begin{pmatrix}
0_{n_{B}\times n_{B}} & I_{n_{B}\times n_{B}}\\
-I_{n_{B}\times n_{B}} & I_{n_{B}\times n_{B}}%
\end{pmatrix}
.
\]
The purity of the corresponding state $\widehat{\rho}$ is given by the formula
(\cite{Birkbis})
\[
\operatorname*{Tr}(\widehat{\rho}^{2})=\left(  \tfrac{\hbar}{2}\right)
^{n}(\det\Sigma)^{-1/2}%
\]
and we have $\operatorname*{Tr}(\widehat{\rho}^{2})=1$ (e.g.. $\widehat{\rho}$
is a pure state) if and only if $\Sigma=\frac{\hbar}{2}S^{T}S$ for some
symplectic matrix $S$ (i.e. $S^{T}JS=J$). We will write the covariance matrix
as
\begin{equation}
\Sigma=%
\begin{pmatrix}
\Sigma_{AA} & \Sigma_{AB}\\
\Sigma_{BA} & \Sigma_{BB}%
\end{pmatrix}
\text{ with\textit{ }}\Sigma_{BA}=\Sigma_{AB}^{T} \label{sigmab}%
\end{equation}
the blocks $\Sigma_{AA}$, $\Sigma_{AB}$, $\Sigma_{BA}$, $\Sigma_{BB}$ having
dimensions $2n_{A}\times2n_{A}$, $2n_{A}\times2n_{B}$, $2n_{B}\times2n_{A}$,
$2n_{B}\times2n_{B}$, respectively. A calculation involving Gaussian integrals
\cite{JMP,DiGoPra} shows that the partial trace $\widehat{\rho}^{A}$ is a
Gaussian state with Wigner distribution%
\begin{equation}
\rho_{A}(z_{A})=\frac{1}{(2\pi)^{n_{A}}\sqrt{\det\Sigma_{AA}}}e^{-\frac
{1}{\hbar}z_{A}^{T}(\Sigma_{AA}^{-1})z_{A}}~. \label{rhoaza}%
\end{equation}
This can easily be seen by reducing the problem to the case where the
covariance matrix (\ref{sigmab}) is block diagonal%
\[%
\begin{pmatrix}
\Sigma_{AA} & \Sigma_{AB}\\
\Sigma_{BA} & \Sigma_{BB}%
\end{pmatrix}
=%
\begin{pmatrix}
H^{T} & 0\\
0 & K^{T}%
\end{pmatrix}%
\begin{pmatrix}
\Sigma_{AA}^{\prime} & 0\\
0 & \Sigma_{BB}^{\prime}%
\end{pmatrix}%
\begin{pmatrix}
H & 0\\
0 & K
\end{pmatrix}
\]
with $H\in O(2n_{A})$, $K\in O(2n_{B})$. In particular we have%
\[
\operatorname*{Tr}[(\widehat{\rho}^{A})^{2}]=\left(  \tfrac{\hbar}{2}\right)
^{n}(\det\Sigma_{AA})^{-1/2}%
\]
hence $\widehat{\rho}_{A}$ is a pure state if and only if $\det\Sigma
_{AA}=(\hbar/2)^{2n}$; this is the case if and only if $\Sigma_{AA}%
=\frac{\hbar}{2}S_{A}^{T}S_{A}$ where $S_{A}$ is a symplectic matrix in
$\mathbb{R}^{2n_{A}}$.

\section{Purification}

We recall the following: let $\widehat{\rho}^{A}\in\mathcal{L}_{1}%
(L^{2}(\mathbb{R}^{n_{A}}))$ be a density matrix of rank $r\leq\infty$. Then
there exists a Hilbert space $H_{B}$ with $\dim H_{B}=r$ and a pure state
$|\psi\rangle$ such that $|\psi\rangle\langle\psi|\in\mathcal{L}_{1}%
(L^{2}(\mathbb{R}^{n_{A}})\otimes H_{B})$ and $\widehat{\rho}^{A}%
=\operatorname*{Tr}\nolimits^{B}(|\psi\rangle\langle\psi|)$ (we assume
throughout that $W\psi$ satisfies the conditions of Lemma \ref{Lemma1}). The
state $|\psi\rangle\langle\psi|$, which is not uniquely defined, is called a
\emph{purification} of $\widehat{\rho}^{A}$. In what follows we assume that
$\widehat{\rho}^{A}$ has maximal rank $r=\infty$ and we identify $H_{B}$ with
$L^{2}(\mathbb{R}^{n_{B}})$. The purification is constructed as follows: one
starts with the spectral decomposition
\[
\widehat{\rho}^{A}=\sum_{j}\lambda_{j}|\phi_{j}^{A}\rangle\langle\phi_{j}%
^{A}|\text{ \ },\text{ \ }\sum_{j}\lambda_{j}=1\text{ \ },\text{ \ }%
\lambda_{j}\geq\geq0
\]
and one then defines $\psi\in L^{2}(\mathbb{R}^{n_{A}})\otimes L^{2}%
(\mathbb{R}^{n_{B}})$ by its Schmidt decomposition \cite{Nielsen}%
\begin{equation}
\psi=\sum_{j}\sqrt{\lambda_{j}}\phi_{j}^{A}\otimes\phi_{j}^{B}\text{ \ ,
\ }\sum_{j}\lambda_{j}=\langle\psi|\psi\rangle=1~. \label{Schmidt1}%
\end{equation}
We have
\begin{equation}
\widehat{\rho}^{A}=\operatorname*{Tr}\nolimits^{B}(|\psi\rangle\langle
\psi|)=\sum_{j}\lambda_{j}|\phi_{j}^{A}\rangle\langle\phi_{j}^{A}|
\label{rhoabis}%
\end{equation}
so, by Theorem \ref{Thm1}, the Wigner distribution $\rho^{A}(z_{A})$ of
$\widehat{\rho}^{A}$ is
\begin{equation}
\int_{\mathbb{R}^{2n_{B}}}W\psi(z_{A},z_{B})dz_{B}=\sum_{j}\lambda_{j}%
W^{A}\phi_{j}^{A}(z_{A}) \label{Wigsum}%
\end{equation}
which is formula (\ref{F2}).

\section{Concluding remarks}

The main result of this work is the partial trace formula (\ref{F1}) in
Theorem \ref{Thm1}. While this formula is hardly surprising (and seems to be
part of the \textquotedblleft folklore\textquotedblright\ in physics), its
proof is not trivial, and we have been unable to find a rigorous treatment
thereof in the literature (mathematical or physical); mentions of this result
can be found in \cite{R1,R2,R3}, however no proof whatsoever is given in these
references. This result is particularly important when one uses methods from
harmonic analysis in phase space to study problems in quantum mechanics. From
our point of view, this approach is more fruitful than abstract operator
theory because it leads to quantities that can be explicitly calculated
(either directly, or using numerical algorithms). Theorem \ref{Thm1} and its
consequences (as shortly outlined in the rest of the article) could be
instrumental in advancing the study of entanglement and entropy of partial
states because it allows direct explicit calculations involving the Weyl
expression of these partial states (in \cite{DiGoPra} we used formula
(\ref{F1}) without proof to study the separability of Gaussian States)..

\begin{acknowledgement}
Maurice de Gosson has been financed by the Grant P 33447-N of the Austrian
Science Fund FWF.
\end{acknowledgement}

There has been no data generated or analyzed in his work. Neither is there any
conflict of interest.

\end{document}